\documentclass{iopart}

\newcommand{\beeq}{\begin{equation}}
\newcommand{\eneq}{\end{equation}}
\newcommand{\beqn}{\begin{eqnarray}}
\newcommand{\eeqn}{\end{eqnarray}}

\def\dd{\partial}
\def\la{\raise.16ex\hbox{$\langle$} \, }
\def\ra{\, \raise.16ex\hbox{$\rangle$} }
\def\ran{\raise.16ex\hbox{$\rangle$} }
\def\go{\rightarrow}

\def\next{{~~~,~~~}}
\def\onehalf{ \hbox{${1\over 2}$} }

\def\psibar{ \psi \kern-.65em\raise.6em\hbox{$-$}\lower.6em\hbox{} }
\def\L{ {\cal L} }

\begin{document}

{\small \noindent
9 September 1998 \hfill UMN-TH-1717/98 
\rightline{hep-th/9809066}
}

\title[S=1/2 Heisenberg chain and the Schwinger model]{Antiferromagnetic 
$S=\onehalf$ Heisenberg Chain and the Two-flavor Massless Schwinger Model
}

\author{Yutaka Hosotani}

\address{School of Physics and Astronomy, University of Minnesota,
Minneapolis, MN 55455, USA}

\begin{abstract}
An antiferromagnetic $S={1\over 2}$ Heisenberg chain is mapped to the 
two-flavor massless Schwinger model at $\theta=\pi$.  The electromagnetic 
coupling constant and velocity of light in the Schwinger model are determined
in terms of the Heisenberg coupling and lattice spacing in the spin chain
system.  
\end{abstract}


\vskip .5cm

\noindent
In the previous paper \cite{Hosotani1} we have shown that an $S={1\over
2}$ spin chain
\beeq
H_{\rm chain}(\vec S) = J \sum \vec S_n \cdot \vec S_{n+1}  
\qquad (J>0)
\label{spinchain1}
\eneq
corresponds to the two-flavor massless Schwinger model (two-dimensional
QED) in a uniform background charge density:
\beeq
\hskip -1.0cm
\L_{\rm QED} (A_\mu, \psi) = - {1\over 4e^2} \, F_{\mu\nu}^2
+ \sum_{a=1}^2 c \, \psibar^{(a)}  \gamma^\mu 
  \Big( i \hbar \dd_\mu - {1\over c} A_\mu \Big)
\psi^{(a)} + {1\over \ell} A_0  ~.
\label{Schwinger1}
\eneq
Here $\ell$ is the lattice spacing in the spin chain model.  The coupling
constant $e$ and velocity of light $c$ are related to $J$ and $\ell$ by
$e = k \sqrt{J/\ell}$ and $c = 2\ell J/\pi\hbar$, where $k$ is a constant
of  O(1)  left undetermined.  The Schwinger model contains one
more parameter.  Its ground state is the $\theta$ vacuum.   The value of
$\theta$
 also remained  undetermined. In this Letter we shall show that
$k \sim 2.225$ and $\theta=\pi$ with the aid of the result from the Bethe
ansatz.\cite{Bethe}

The correspondence between the spin chain and Schwinger model has been
noticed by many authors.\cite{Wiegmann}   Deductive derivation was achieved
in ref.\ \cite{Hosotani1}.The original spin degree is
tranformed to the flavor index of the Dirac field $\psi$, whereas the 
odd-even site index of the lattice becomes the spin degree of the Dirac
field.  The finiteness of the lattice spacing $\ell$ must be taken into
account in establishing the correspondence.  In the naive continuum limit
$\ell \go 0$ with $c$ kept fixed, the charge $e$ diverges
as $1/\ell$ and all massive modes of the Schwinger model decouple.
These massive modes play an important role when the correspondence is
applied to the spin ladder problem.

The determination of the charge $e$, or the
parameter $k$, was attempted in \cite{Ichinose}  following the
argument of \cite{Hosotani1}.  More recently Berruto \etal have examined the
correspondence between the spin chain model and the two-flavor lattice
Schwinger model and have found that both model have the  low energy
excitation spectrum of the same pattern.\cite{Grignani} 

To determine $k$ and $\theta$, first notice that the Schwinger model
(\ref{Schwinger1}) is bosonized in terms of two scalar fields.\cite{Hosotani2,
Hosotani3} 
One of them  remains massless, which corresponds to the gapless
excitation in the Bethe ansatz solution.
The other has a mass $\mu$ where
$\mu c^2 = (2e^2 \hbar c/\pi)^{1/2}$.  In the two-flavor massless
Schwinger model the $SU(2)$ chiral condensate $\la \psibar^{(a)}
\psi^{(b)}\ra_\theta$ vanishes. However the $U(1)$ chiral condensate
is nonvanishing: \cite{Hosotani3} 
\beqn
\hskip -1.5cm
\la \psibar^{(1)}\psi^{(1)} \psibar^{(2)}\psi^{(2)} \ra_\theta
&=& \la \psi_L^{(1)\dagger} \psi_L^{(2)\dagger} 
   \psi_R^{(2)} \psi_R^{(1)} \ra_\theta 
+ \la \psi_R^{(1)\dagger} \psi_R^{(2)\dagger} 
   \psi_L^{(2)} \psi_L^{(1)} \ra_\theta \cr
\noalign{\kern 5pt}
&=& \Big( {e^\gamma\over 4\pi} \Big)^2 
     \Big( {\mu c\over \hbar} \Big)^2 (e^{i\theta} + e^{-i\theta})  \cr
\noalign{\kern 5pt}
&=&  {e^{2\gamma} k^2\over 8\pi^2 \ell^2}  ~ \cos\theta ~~.
\label{condensate}
\eeqn
 
On the other hand the scalar density operator $\psibar^{(a)} \psi^{(a)}$
is related to the spin operator.\cite{Hosotani1} Recall that when
a spin operator is written as $\vec S_n = c^\dagger_n\onehalf \vec \sigma
c^{}_n$,
\beeq
\cases{\sqrt{2\ell} \, \psi_1^{(a)} 
  = (-i)^{2s-1} ~ c_{2s-1,a} &at odd site\cr
\sqrt{2\ell} \, \psi_2^{(a)}
  = ~(-i)^{2s}~~ c_{2s,a}&at even site\cr}
\label{correspondence1}
\eneq
in the representation $(\gamma^0, \gamma^1, \gamma^5)=(\sigma_3, i\sigma_2,
\sigma_1)$. With the half-filling condition $c^\dagger_n c_n =1$
\beeq
\pmatrix{ \psibar^{(1)} \psi^{(1)} \cr \psibar^{(2)} \psi^{(2)} \cr}
= \pm ~ {1\over 2\ell} ~ ( S^z_{2s-1} - S^z_{2s}) ~~.
\label{correspondence2}
\eneq
Hence
\beqn
\la \psibar^{(1)}\psi^{(1)} \psibar^{(2)}\psi^{(2)} \ra_\theta
&=&- {1\over 4\ell^2} \la ( S^z_{2s-1} - S^z_{2s})^2 \ra \cr
\noalign{\kern 5pt}
&=& - {1\over 6\ell^2} \Big\{ {3\over 4} - \la \vec S_n \cdot \vec S_{n+1} \ra
\Big\}   \cr
\noalign{\kern 5pt}
&=& - {1\over 6\ell^2} \Big\{ {1\over 2} + \ln 2 \Big\} ~.
\label{correspondence3}
\eeqn
In the last equality we have made use of the exact result from the Bethe ansatz
$\la \vec S_n \cdot \vec S_{n+1} \ra = - \ln 2 + {1\over 4}$. \cite{Bethe} 

We equate  (\ref{condensate}) and (\ref{correspondence3}).  The original
spin chain model is parity invariant, which implies that  the Schwinger model
with  only $\theta=0$ or $\pi$ corresponds to the spin chain.  As 
(\ref{correspondence3}) is negative, 
\beeq
\theta=\pi \next k = {\pi\over e^\gamma} \sqrt{ {2\over 3} + {4\over 3} \ln 2
}  =2.225 ~~.
\label{k-value}
\eneq

This is the result of the paper.   The excitation energy of the massive
mode in the  Schwinger model is $\mu c^2 = 2kJ/\pi = 1.416 \, J$, whereas the
Compton wave length is $\hbar/\mu c = \ell/k = .449 \, \ell$.

In ref.\ \cite{Hosotani1} this correspondence was applied to the two-leg
spin ladder system.  In experimental samples $J'\sim J$ and the spin
gap $\Delta_{\rm spin} \sim .5 J'$.\cite{exp}  It was found   that the
spin gap is given by
$\Delta_{\rm spin} = 0.25 k |J'|$ where $J'$ is the interchain
Heisenberg coupling  when $|J'| \ll J$.  With the $k$ value substituted
this yields $0.556 |J'|$, which should be compared with the Monte
Carlo result $0.41 |J'|$.\cite{numerical}   The discrepancy is attributed
to the approximation employed in the spin ladder problem. 

When $J'=0$, there appear two gapless modes, say $\chi_1$ and $\chi_2$,
associated with two chains. When $J' \not= 0$, these gapless modes
acquire gaps.  In general two combinations $\chi_\pm =\chi_1 \pm \chi_2$
are expected to  acquire different gaps.  However, in the
approximation scheme used in ref.\ \cite{Hosotani1}, $\chi_+$ and $\chi_-$ 
remain degenerate.  Shelton \etal have examined the two-leg spin ladder
in a slightly different approximation scheme and have found that
the two modes acquire different gaps, though they could not determine
their absolute magnitude.\cite{Shelton}    It is a challenging problem to
improve the approximation to find more accurate excitation spectrum
in the spin ladder system.

\section*{Acknowledgements}
I would like to thank Paul Wiegmann and Manfred Sigrist for  
helpful discussions, and the Yukawa Institute  for Theoretical
Physics for its hospitality where most of this work was done. This work
was supported in part  by the U.S.\ Department of Energy under contracts
DE-FG02-94ER-40823.

\def\jnl#1#2#3#4{#4 {#1}{\bf #2}  #3}
\def\Zphys{{\em Z.\ Phys.} }
\def\jssc{{\em J.\ Solid State Chem.\ }}
\def\jpsJ{{\em J.\ Phys.\ Soc.\ Japan }}
\def\ptps{{\em Prog.\ Theoret.\ Phys.\ Suppl.\ }}

\def\myshift{ \hglue -.35cm}

\end{document}